\documentclass[aps,prl,twocolumn,showpacs,superscriptaddress]{revtex4}

\usepackage{graphicx}
\usepackage{dcolumn}
\usepackage{bm}
\usepackage{stmaryrd}
\usepackage{latexsym}
\usepackage{amssymb}
\usepackage{amsfonts}
\usepackage{amsmath}
\usepackage{fancybox}
\usepackage{verbatim}
\usepackage{ulem}

\begin{document}

\title{Irreversibility in response to forces acting on graphene
sheets}

\author{N. Abedpour}
\affiliation{Department of Physics, Sharif University of Technology, 11365-9161,
Tehran, Iran}
\affiliation{School of Physics, Institute for Research in Fundamental Sciences, (IPM) Tehran 19395-5531, Iran}
\author{Reza Asgari}
\affiliation{School of Physics, Institute for Research in Fundamental Sciences, (IPM) Tehran 19395-5531, Iran}
\author{M. Reza Rahimi Tabar}
\affiliation{Department of Physics, Sharif University of Technology, 11365-9161,
Tehran, Iran}
\affiliation{Fachbereich Physik, Universit\"{a}t Osnabr\"{u}ck, Barbarastra{\ss}, 49076 Osnabr\"{u}ck, Germany}

\begin{abstract}

The amount of rippling in graphene sheets is related to the
interactions with the substrate or with the suspending structure.
Here, we report on an irreversibility in the response to forces that
act on suspended graphene sheets. This may explain why one always
observes a ripple structure on suspended graphene. We show that a
compression-relaxation mechanism produces static ripples on graphene
sheets and determine a peculiar temperature $T_c$, such that for
$T<T_c$ the free-energy of the rippled graphene is smaller than that
of roughened graphene. We also show that $T_c$ depends on the
structural parameters and increases with increasing sample size.

\end{abstract}

\maketitle

\noindent { Introduction}--- Two-dimensional graphene crystals
\cite{novoselov} have attracted considerable attention both
experimentally and theoretically, due to their unusual electronic
properties \cite{yafis}. Ripples or undulations were first observed
in freely-suspended graphene flakes in experiments \cite{meyer}. The
ripples affect the electronic properties \cite{juan}, such as the
conductivity and the quantum transport properties. Charge
inhomogeneities due to the ripples have also been observed in
graphene \cite{inhomo}. A new method to fabricate periodically
rippled graphene on Ru(0001) under ultrahigh vacuum conditions was
reported in \cite{parga}. The ripples in graphene were studied
theoretically \cite{flagg}, and it was claimed that the ripples may
be explained as a consequence of absorbed molecules sitting on
random sites.

There are two points of view on the physics of the ripples. Meyer
{\it et al.} \cite{meyer} proposed that the reproducible appearance
of the ripples across samples indicates that it is an intrinsic
effect. They emphasized that the homogeneity and isotropy of the
ripples are not compatible with the assumption of an incompressible
sheet. They estimated a local strain of up to 1\% for a single-layer
flake. Using the generic free-energy for the long wavelength
deformations \cite{nelson}, Castro Neto and Kim \cite{castro}, on
the other hand, argued that graphene may be considered as an atomic
thin membrane, with its physics being also similar to a soft
membrane. In this point of view the ripples are not intrinsic and
can be the results of the environment, such as the substrate in the
system.

Recently, direct observation and controlled creation of periodic
ripples in suspended graphene sheet was reported \cite{bao} by using
both spontaneously and thermally generated strains via varying the
substrate and annealing conditions. The experimental measurements
indicated that the ripples were induced by the preexisting
longitudinal strains in graphene. It was observed at temperatures
about 500 K that graphene sheets are flat, however, upon cooling
down to room temperature, the ripples invariably appear.

In this Letter we report on an irreversibility in response to forces
that act on the suspended graphene sheets that may explain why one
always observes the rippled structure on graphene in experiments. We
show that a compression-relaxation mechanism can produce static
ripples on graphene sheets. We determine a peculiar temperature
$T_c$ such that for temperatures less than $T_c$ the free-energy of
the rippled graphene is smaller than that of roughened graphene. We
also show that for sample with size $400\times 200$ atoms, $T_c
\simeq 90$ K and, moreover, $T_c$ is an increasing function of the
sample size.

\noindent {Theory and Model}--- We used molecular dynamics (MD)
simulations with the empirical interatomic interaction potential
due to Brenner \cite{brenner}, i.e., carbon-carbon interaction in
hydrocarbons that contains three-body interaction. Many-body effects
of electron system, on average, are considered in the Bernner
potential through the bond-order term. We employed the Nos\'e-Hoover
thermostat to control the temperature, when we used a canonical
(NVT) ensemble in the MD simulations. We note that although the
Brenner potential is not entirely a quantum mechanical potential, it
predicts the correct mechanical properties of the structures with
carbon atoms by using the classical MD simulations \cite{bee}.

Here, we show that the compression and then relaxation in one or two
directions [$x$ (arm-chair) and $y$ (zigzag)] of graphene sheet can
produce static ripples, which means that there is an irreversibility
in response to forces acting on segments of the graphene sheets.
Indeed, we have found that if one compress the surface in one
direction, say the $x-$direction (arm-chair), after compressing
about $0.13\% L$, where $L$ is the size of the simulation sample,
the ripples will appear. After the ripples emerge, we move back the
boundaries to their {\it initial} positions. We then observe that
after doing the compression-relaxation processes, the ripples {\it
survive}, hence implying that the compression process is not
reversible. It is worthwhile mentioning that we also simulated
tethered membranes \cite{kantor} and repeated the
compression-relaxation procedure. We obtained no any indication of
irreversibility in response to the forces acting on the membranes.

In the case of graphene, if the compression amount becomes larger
than the critical value (here, $0.13\% L$, which also depends on the
temperature and the size of simulation sample), the graphene sheet
bends and, therefore, no ripple appears. The typical height variance
of the rippled graphene is about $5\;\AA$ at $T=50$ K. Our
simulation results show that the wavelength of the static ripples do
not change with the size of the sample. Moreover, we observe that
the surface roughness (the variance of the height fluctuations) does
not change after the relaxation and, therefore, the ripples are
static. Thus, we might state that any primary stress on graphene
sheet, for example in its preparation in the experiments, can {\it
construct} ripples that will {\it survive} during the experimental
measurements (see, for example, \cite{boukhvalov}).

Let us first determine the average wavelength of the ripples after
relaxing the system. To do so we calculate the two-dimensional
Fourier components of the height-height correlations, $G(|{\bf q}|)=
<\left|h({\bf q})\right|^2>$. Figure 1 shows $G({|\bf q|})$ as a
function of ${|\bf q|}/q_0$, in logarithmic scales, for both the
roughened (no ripples) and relaxed states in which we have stable
ripples. Here $q_0=2\pi/L$, with $L$ being the length of graphene in
the $x-$direction. In the inset of Fig. 1 the same plot in linear
scale for ${|\bf q|}$ is shown to clarify a peak around ${|\bf
q|}\simeq 10 q_0$ that corresponds to about $85\;\AA$ at 50 K. This
is the average wavelength of the ripples and is near to the value
observed experimentally \cite{meyer} and calculated numerically
\cite{Faso,nima}. In addition, one can derive the wavelength of the
ripples by calculating the first minimum of the second moments of
the height increments fluctuations $< |h(x_1) - h(x_2)|^2> $ with
respect to relative distance, $|x_1 - x_2|$.

The scale-dependence of $<\left|h({\bf q})\right|^2>$ is
proportional to $1/{|\bf q|}^\alpha$, where $\alpha\simeq 4$ at
temperature 50 K. Consequently, our results predict that the bending
rigidity term prevails with respect to the surface tension in
graphene at short distances \cite{safran}. Note that the
contribution of surface tension is a term like $T/\sigma {|\bf
q|}^2$, whereas the contribution of the bending rigidity is
$T/\kappa {|\bf q|}^4$, where $\sigma$ and $\kappa$ are the
interfacial tension and bending modulus, respectively \cite{safran}.
The exponent $\alpha$ might generally be smaller than 4, due to
thermal fluctuations, surface tension and anharmonic corrections
\cite{nelson}. We estimate $\kappa$, the bending rigidity or bending
modulus, using the relation, $\kappa^{-1} \simeq |{\bf q}|^4 <
|h({\bf q})|^2>/Nk_BT$. Plotting $\kappa$ vs $|{\bf q}|/q_0$ shows
that the bending rigidity is almost constant for $20<|{\bf
q}|/q_0<100$, with $\kappa \simeq 1$ eV$^{-1}$.

Thus, for a given temperature, the compression-relaxation mechanism
produces the ripples, and graphene has at least two "states"
simultaneously, the "rough" or normal sheet (no ripple) and the
"rippled" structure. The question now is, which state is more
stable? To answer this question one should calculate the free
energies of the two states and determine which state has a smaller
free-energy. In what follows we calculate the free-energy difference
of the rippled and roughened states of graphene sheets.

To compute the free-energy, we employed a well-known method (c.f.,
Haile \cite{allen}) in which one defines a continuous variable
$\lambda$ for distinguishing two different states \cite{Jarzynski}.
Suppose that by varying an external parameter, such as slow
compression and relaxation of the graphene, the system can go from
an initial state {\it i} (rough) to a final state {\it f} (rippled).
When the parameters are changed infinitely slowly along some path
from {\it i} to {\it f} in the parameter space, then the total work
$W$ performed on the system is equal to the Helmholtz free-energy
difference between the initial and final configurations. In
contrast, when the parameters are switched along the path at a
finite rate, Jarzynski found that \cite{ Jarzynski}:
\begin{equation}
\Delta A =  -\frac{1}{\beta} \ln {\overline{{\exp(-\beta W)}}}
\end{equation}
where overbar denotes an average over an ensemble of measurements of
$W$. We ran the MD simulation to very long times in order to slowly
pass the intermediate quasistable states. In practice, for every
step of compression-relaxation, we checked whether the system was in
equilibrium. We ensured the existence of the true equilibrium
condition by checking the stability of the internal-energy
fluctuations. Eventually, the problem of calculating $\Delta A$ is
the same as calculating the averaged $W$.

\begin{figure}[t]
\begin{center}
\includegraphics[width=0.8\linewidth]{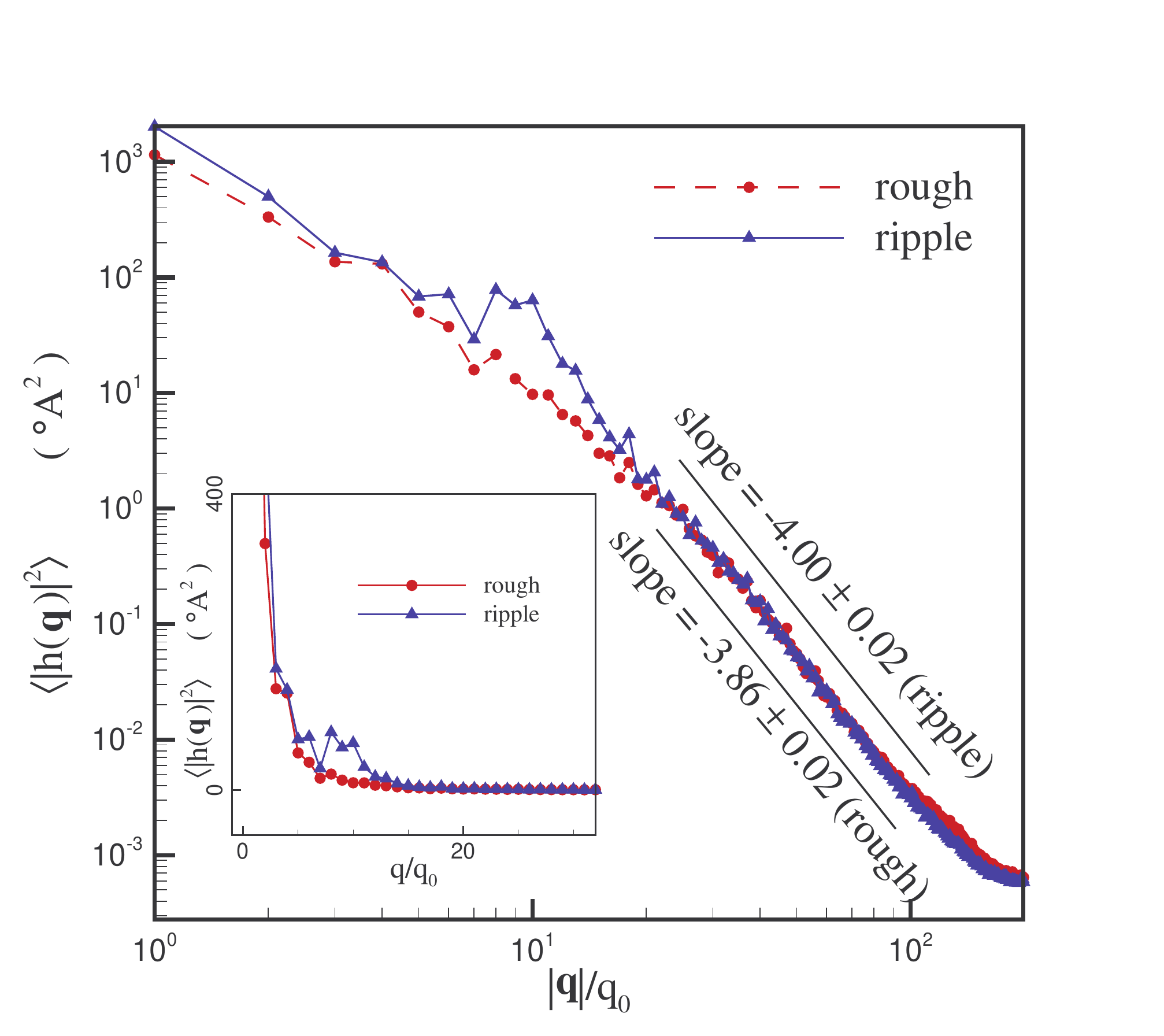}
\caption{(Color online) Log-log plot of $<|h({\bf q})|^2>$ as a
function of ${\bf q}/q_0$ (log scaled) both in the rough and ripple
cases.  In the inset, $<|h({\bf q})|^2>$ is shown as a function of
$|{\bf q}|/q_0$ to clarify a peak around $|{\bf q}|\simeq 10 q_0$. Graphene sheet incorporates $80000$ atoms at $50 K$. }
\end{center}
\end{figure}

\begin{figure}[h]
\begin{center}
\includegraphics[width=0.8\linewidth]{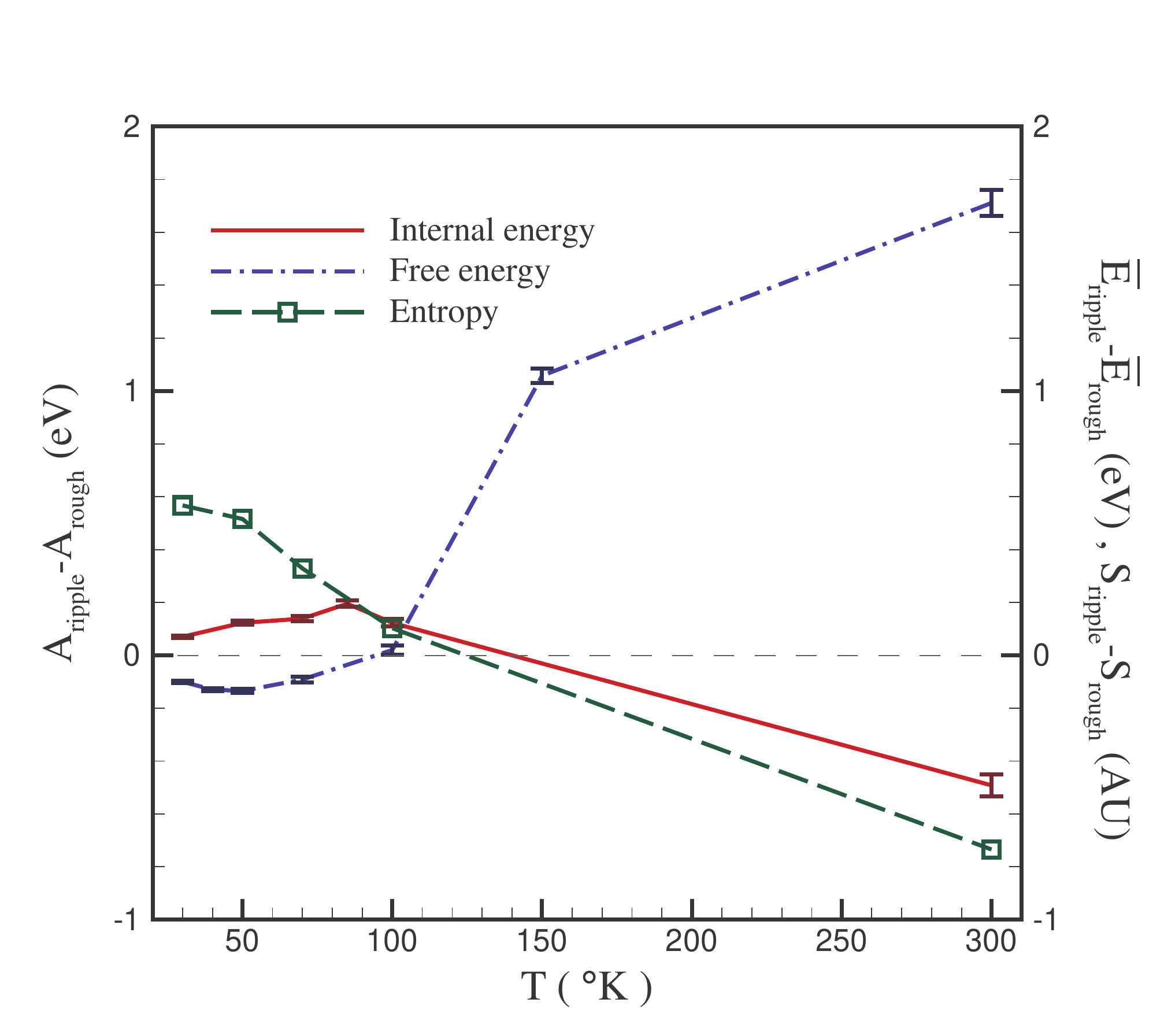}
\end{center}
\caption{\footnotesize{(Color online) The dependence of the free-energy and internal-energy (entropy) differences as a function of
temperature.}}
\end{figure}

\begin{figure}[h]
\begin{center}
\includegraphics[width=0.85\linewidth]{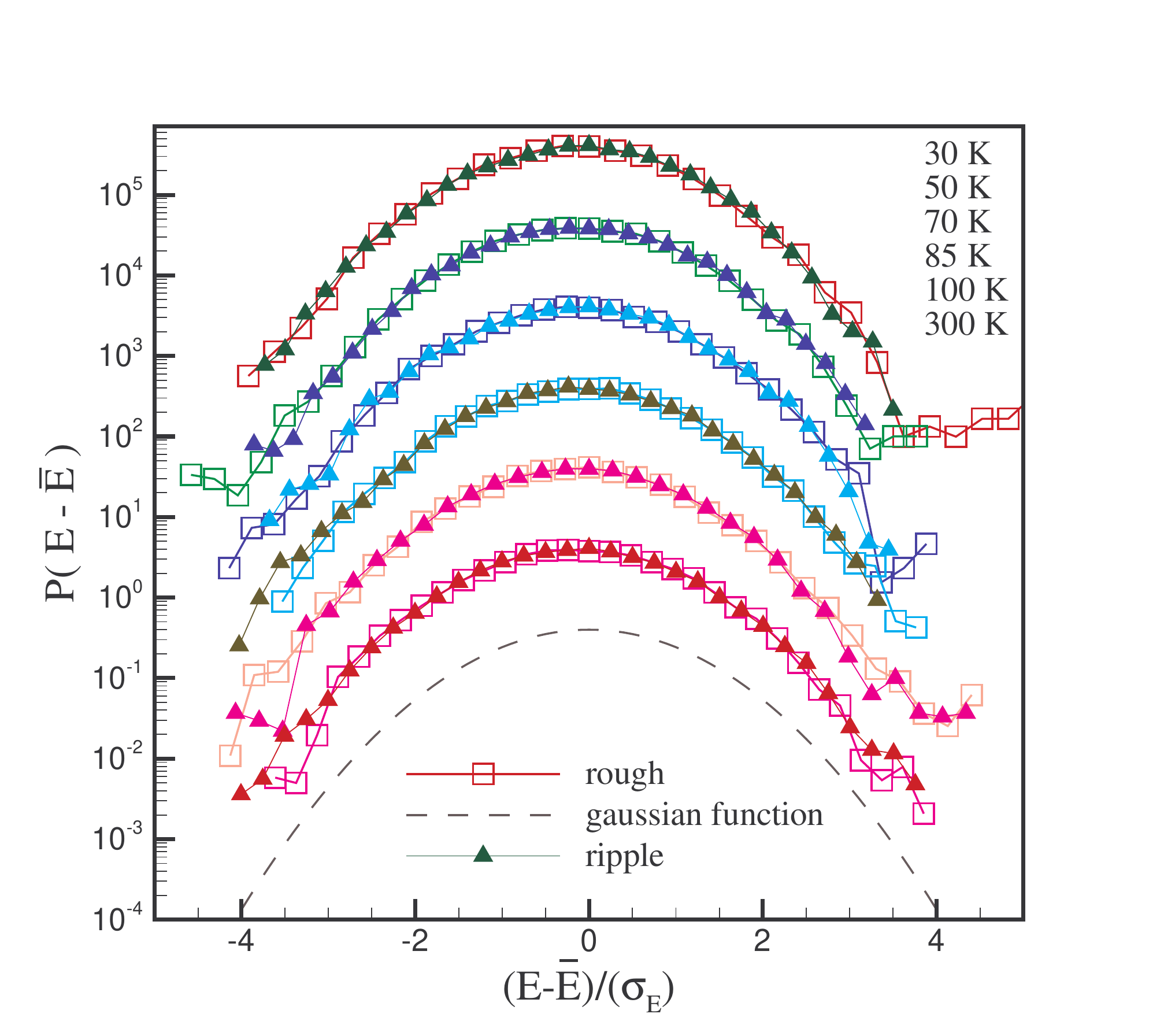}
\end{center}
\caption{\footnotesize{(Color online) Probability distribution
function of the total energy $E$ for the rippled and rough states
for $T =30, 50, 70, 85, 100$ and 300 K (from top to bottom). The dashed curves are the
Gaussian probability distribution function.  For clarity, the PDFs
were shifted upward. }}
\end{figure}

In Fig. 2 the free-energy differences $A_{\rm ripple}-A_{\rm rough}$
is given as a function of $T$. We used the numerical results for a
graphene sheet incorporating $N=80000$ atoms $(400\times 200)$ at
various temperatures. It appears that the ripples are stable at low
temperatures, namely, below $T_c\approx 90$ K, such that above $T_c$
the rough state is more stable. This feature is in good agreement
with recent experimental observation \cite{bao}. Here, we would like
to point out that the potential energy of the carbon-carbon
interaction in the compression process is different from that in the
relaxation process since the relative positions of the atoms in
these two configurations are different. Note that the morphology of
the surface depends strongly to the potential energy.

 We also tested
the dependence of $T_c$ on the size of the samples by simulating the
systems with $600\times 200$ and $800\times 200$ atoms, and found
their characteristic temperatures to be $T_c\simeq 115$ and
$T_c\simeq 140$, respectively. Moreover, we found that the
wavelength of the ripples depends on $T$ as $\lambda\simeq 35\ln(T)
- 55$, but it does not depend on the system size. Accordingly, we
calculated the entropy difference of the two states and showed that
for temperatures less than $T_c$, the rippled state has a higher
entropy and is stable. We plotted the internal energy difference of
the two states is shown in Fig. 2.
 The value of $T_c$ can be also determined from the local
stored stress on a graphene sheet; we will report the results
elsewhere \cite{mehdi}.

We may expect that similar to second-order phase transitions the
probability distribution function (PDF) of the total internal energy
possesses different shapes for rough and ripple states, and exhibit
non-Gaussian behavior. In Fig. 3 the PDF of the total internal
energy $E$ for the ripple and rough states are presented for $T=30,
50, 70, 85, 100$ and 300 K. To calculate the PDF, we used 200
ensembles of roughened and rippled graphenes, incorporating $ 400
\times 200$ atoms. We observed that the PDF has a Gaussian form for
both states indicating that there is no longer second--order phase
transition in the system. We also checked the Gaussian nature of the
PDF by using the $\chi^2$ test \cite{chi}. The dashed curves
represent the Gaussian PDF.

As we argued earlier, there are at least two states for graphene
sheets for a given temperature. A question raised is,
whether or not, there is any possibility of a transition from one state
to another? One possible way for such a transition with fixed
graphene sheet size is to increase the temperature. For this purpose
we simulated the graphene sheet with $80 \times 40$ atoms and, after
carrying out the compression-relaxation process, the ripple
structure appeared at $T=55$ K (see the upper figure of Fig. 4). We
then increased the temperature very slowly. As shown in Fig. 4, the
ripples begin to disappear at high temperatures. At $T=55$ K, we
have almost two wavelength of the ripples; however, at higher
temperatures there is one wavelength at $T=320$ K, and finally one
half of the wavelength at $T=520$ K remains. The final step may be
called rough state. Accordingly, the energy barrier of two states
may be estimated by $465 k_B =0.04$ eV (or $=0.0125$ meV per
particle) for a sample with 3200 atoms, where $k_B$ is the
Boltzmann's constant. Such a transition has been observed
experimentally in [10]. They argued that the disappearing of ripples
in high temperature is due to the fact that graphene has negative
thermal expansion coefficient.

As mentioned above, for temperatures less than $T_c$, the free-energy of the rippled state is smaller than that of the free-energy
of the roughened graphene sheet. However, there is a possibility of
having a transition from the rippled state to the roughened state,
due to a tunneling-type transition. To detect this transition, we
checked that the height fluctuations variance of the rippled
graphene sheet is stable with time. A sample size of $80\times 40$
atoms was used at $T=55$ K. The simulations showed that there is no
transition from the rippled to roughened state at a constant
temperature $T$ less than $T_c$, at least up to available time
scales in the MD simulations. The probability for such transition is
about $ \exp(- 465/T)$ for a sample with size 3200 atoms. We remind
that the variance of height fluctuations in graphene are about $5
\;\AA$ and $1$ nm, for rough and rippled states, respectively.

\begin{figure}[h]
\begin{center}
\includegraphics[width=0.95\linewidth]{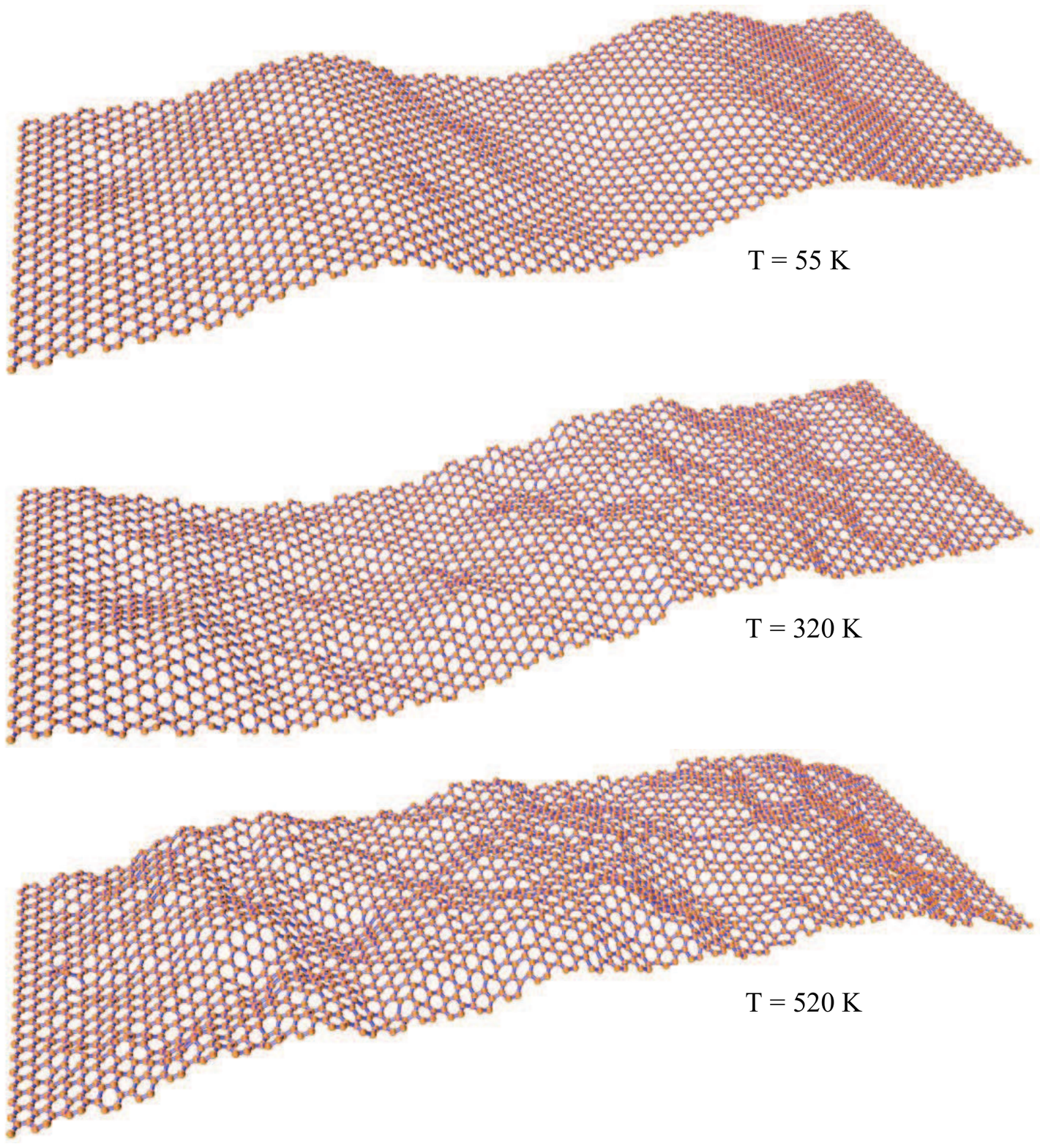}
\caption{\footnotesize{(Color online) Transition from rippled state
(upper snapshot) to the rough state, due to the increasing of the
temperature from $55 K$ to $520 K$.}}\label{lambda}
\end{center}
\end{figure}

In summary, we have used a compression-relaxation mechanism to
produce rippled structures on graphene sheets. The constructed
ripples {\it survive} even though the system is relaxed to its
initial position. In the closed-path loop, we calculated the total
work and, hence, the free-energy difference of the rippled and
roughened states. Our numerical results show that for sample with
$400 \times 200$ atoms and below $T_c \approx 90$ K, the rippled
surface is stable and the entropy of the ripples should be larger
than that of the rough state. However, above $T_c$ the rough state
is more stable. The rippled and rough structures are also related to
the morphology of such systems and we, therefore, expect that the
our simulations yield the correct and
 new results for the free-energy of the rippled and
roughened graphene. We have done similar simulations for a bilayer
graphene and observed that, for a given temperature, the wavelength
of the static ripples are larger than that for a monolayer graphene.
We will report the results for bilayer graphene elsewhere.

 \noindent { Acknowledgments}--- We
thank A.K. Geim, , M. I. Katsnelson, P. Maa{\ss}, A. H. MacDonald
and M. Sahimi for very important comments and discussions. We also
thank M. Neek-Amal for early contributions to the numerical work.

\end{document}